\documentstyle[12pt]{article}
\begin{document}
\newcommand{\siml}{\stackrel{<}{\sim}}
\newcommand{\simg}{\stackrel{>}{\sim}}
\newcommand{\lleq}{\stackrel{<}{=}}

\baselineskip=1.333\baselineskip
\baselineskip=0.8\baselineskip

\noindent
\begin{center}
{\Large\bf
Non-extensive thermodynamics of \\
transition-metal nanoclusters
\footnote{To be submitted to Prog. Mat. Sci.:
Festschrift Proceedings for David's 60th}
} 
\end{center}

\begin{center}
Hideo Hasegawa
\footnote{E-mail:  hasegawa@u-gakugei.ac.jp}
\end{center}

\begin{center}
{\it Department of Physics, Tokyo Gakugei University  \\
Koganei, Tokyo 184-8501, Japan}
\end{center}
\begin{center}
({\today})
\end{center}
\thispagestyle{myheadings}

\noindent
{\bf Abstract}   

In recent years, much study has been made by applying
the {\it non-extensive statistics} (NES) to
various non-extensive systems 
where the entropy and/or energy
are not necessarily proportional to
the number of their constituent subsystems.
The non-extensivity may be realized 
in many systems
such as physical, chemical and biological ones, 
and also in small-scale nanosystems.

After briefly reviewing the recent development 
in nanomagnetism and the NES,
I have discussed, in this article, NES calculations
of thermodynamical properties of a nanocluster
containing noninteracting $M$ dimers.
With bearing in mind a transition-metal nanocluster,
each of the dimers is assume to be described by
the two-site Hubbard model ({\it a Hubbard dimer}).
The temperature and magnetic-field
dependences of the specific heat, magnetization and
susceptibility have been calculated
by changing $M=1$, 2, 3 and $\infty$, results
for $M=\infty$ corresponding to those of
the conventional Boltzman-Gibbs statistics (BGS).
It has been shown that the thermodynamical 
property of nanoclusters 
containing a small number of dimers
is considerably different from that
of macroscopic counterparts
calculated within the BGS
The specific heat and susceptibility
of spin dimers described by the Heisenberg model
have been discussed also by employing the NES.

%
%

\vspace{0.5cm}

\noindent
{\bf Contents}

\noindent
1. Introduction

1.1 A brief review of nanomagnetism

1.2 Non-extensive statistics

\vspace{0.5cm}

\noindent
2. Non-extensive thermodynamics of Hubbard dimers

2.1 Energy and entropy

2.2 Specific heat

2.3 Magnetization 

2.4 Susceptibility
\vspace{0.5cm}

\noindent
3. Calculated results

3.1 Temperature dependence

3.2 Magnetic-field dependence

\vspace{0.5cm}

\noindent
4. Discussions and conclusions
\vspace{0.5cm}

\noindent
Acknowledgements
\vspace{0.5cm}

\noindent
Appendix. NES for Heisenberg dimers
\vspace{0.5cm}

\noindent
References

\vspace{0.5cm}

\section{Introduction}
\subsection{A brief review of nanomagnetism}

In the last decade, there has been a considerable
interest in atomic engineering, which makes it possible
to create small-scale materials 
with the use of various methods
(for reviews, see Refs. \cite{Bader02,Kach03,Luban04}).
Small-scale magnetic systems ranging from grains (micros),
nanosystems, molecular magnets and atomic clusters,
display a variety of interesting properties.
Magnetic nanosystems consist of small clusters
of magnetic ions embedded within nonmagnetic
ligands or on nonmagnetic substrates.
Nanomagnetism shows interesting properties 
different from bulk magnetism.
Nanoclusters consisting of transition metals such as
${\rm Fe}_N$ ($N$=15-650) \cite{Heer90}, 
${\rm Co}_N$ ($N$=20-200) \cite{Bucher91},
and ${\rm Ni}_N$ ($N$=5-740) \cite{Apsel96} have been synthesized 
by laser vaporization and
their magnetic properties have been measured,
where $N$ denotes the number of atoms per cluster.
Magnitudes of magnetic moments per atom are increased 
with reducing $N$ \cite{Apsel96}.
It is shown that magnetic moments in Co monatomic chains
constructed on Pt substrates
are larger than those in monolayer Co
and bulk Co \cite{Gambardella02}. 
Recently Au nanoparticles with average diameter of 1.9 {\it nm}
(including 212 atoms), which are protected
by polyallyl amine hydrochloride (PAAHC),
are reported to show ferromagnetism while 
bulk Au is diamagnetic \cite{Yamamoto04}.
This is similar to the case of gas-evaporated Pd fine particles
with the average diameter of 11.5 {\it nm}
which show the ferromagnetism whereas bulk Pd is paramagnetic 
\cite{Shinohara03}. 
The magnetic property of four-Ni molecular magnets
with the tetrahedral structure (abbreviated as Ni4) in
metallo-organic substance
$[{\rm Mo}_{12}{\rm O}_{30}(\mu_2-{\rm OH})_{10}
{\rm H}_2\{{\rm Ni}({\rm H}_2{\rm O}_3) \}_4] 
\cdot 14 {\rm H}_2{\rm O}$ has been studied \cite{Postnikov04}.
Their temperature-dependent susceptibility
and magnetization process
have been analyzed by using the Heisenberg model
with the antiferromagnetic 
exchange couplings between Ni atoms \cite{Postnikov04}.
Similar analysis has been made for 
magnetic molecules of Fe$N$  ($N=6,\:8,\:10$ and 12)
\cite{Lasc97}\cite{Gatteshi00},
and V6 \cite{Luban02}.
Extensive studies have been made for
single molecule magnets of Mn12  
in $[{\rm Mn}_{12}{\rm O}_{12}({\rm CH}_3{\rm COO})_{16}
({\rm H}_2{\rm O})_4]$ \cite{Mn12}
and Fe8 in
$[{\rm Fe}_8(tanc)_6 {\rm O}_2 ({\rm OH})_{12}]
{\rm Br}_9 \cdot 9{\rm H}_2 {\rm O}$ \cite{Caciuffo98}.
Both Mn12 and Fe8
behave as large single spins with $S=10$, and
show quantum tunneling of magnetization
and the square-root relaxation,
which are current topics in namomagnetism.
Much attention has been recently paid to 
single molecule magnets
which are either dimers or behave effectively as dimers,
due to their potential use as magnetic storage
and quantum computing.
The iron $S=5/2$ dimer (Fe2) in $[{\rm Fe(OMe)}(dbm)_2]_2$ 
\cite{Fe2}
has a nonmagnetic, singlet ground state and its thermodynamical
property has been analyzed with the use of
the Heisenberg model \cite{Mentrup99}-\cite{Dai03}.
Similar analysis has been made for transition-metal dimers of
V2 \cite{V2}, Cr2 \cite{Cr2},
Co2 \cite{Co2}, Ni2 \cite{Ni2} and Cu2 \cite{Cu2}.

\subsection{Non-extensive statistics}

As the size of systems becomes smaller,
effects of fluctuations and contributions from surface 
play more important roles.
There are currently three approaches to discussing
{\it nanothermodynamics} for small-size systems:
(1) a modification of the Boltzman-Gibbs statistics
(BGS) adding subdivision energy \cite{Hill01},
(2) non-equilibrium thermodynamics including
work fluctuations \cite{Jarz97}, and
(3) the non-extensive statistics (NES)
generalizing the BGS as to take account
of the non-extensive feature of such systems 
\cite{Tsallis88}-\cite{Raja04}.
A comparison between these approaches have been made
in Refs. \cite{Raja04}\cite{Ritort04}.

Before discussing the NES,
let's recall the basic feature of the BGS 
for a system with internal energy $E$ and
entropy $S$, which is immersed in a large reservoir
with energy $E_0$ and entropy $S_0$.
The temperature of the system $T$ is the same as
that of the reservoir $T_0$ where
$T=\delta E/\delta S$ and $T_0 = \delta E_0/\delta S_0$.
If we consider the number of possible microscopic states
of $\Omega (E_0)$ in the reservoir, its entropy
is given by $S_0 = k_B \:{\rm ln} \Omega(E_0)$ where
$k_B$ denotes the Boltzman constant.
The probability of finding the system with the 
energy $E$ is given by 
$p(E)=\Omega(E_0-E)/\Omega(E) \sim {\rm exp}(-E/k_B T)$
with $E \ll E_0$.
When the physical quantity $Q$ 
of a system containing $N$ particles
is expressed by $Q\: \propto N^{\gamma}$,
they are classified into two groups in the BGS:
intensive ($\gamma=0$) and extensive ones ($\gamma=1$).
The temperature and energy are typical
intensive and extensive quantities, respectively.
This is not the case in the NES, as will be shown below.

When a small-scale nanosystem is immersed in a reservoir,
the temperature of the nanosystem is
expected to fluctuate around the temperature of the reservoir $T_0$
because of the smallness of the nanosystem 
and its quasi-thermodynamical equilibrium states.
Then the BGS distribution mentioned above has to be averaged
over the fluctuating temperature.
This idea has been expressed by 
\cite{Wilk00}\cite{Beck02}\cite{Raja04}
\begin{eqnarray}
p(E) &=& \int_0^{\infty} \: d\beta \:e^{-\beta E}\: f^B(\beta) 
\nonumber \\
&=& [1-(1-q)\beta_0 E]^{\frac{1}{1-q}}
\equiv {\rm exp}_q (-\beta_0 \:E),
\end{eqnarray}
with 
\begin{eqnarray}
q&=& 1 + \frac{2}{N}, \\
f^B(\beta)&=& \frac{1}{\Gamma \left( \frac{N}{2} \right)}
\left( \frac{N}{2\beta_0} \right)^{\frac{N}{2}}
\beta^{\frac{N}{2}-1} 
{\rm exp}\left( -\frac{N \beta}{2\beta_0} \right), \\
\beta_0 &=& \frac{1}{k_B T_0}
= \int_0^{\infty} \:d\beta  f(\beta)\:\beta \equiv E(\beta), \\
\frac{2}{N}&=& \frac{E(\beta^2)-E(\beta)^2}{E(\beta)^2},
\end{eqnarray}
where ${\rm exp}_q(x)$ 
denotes the $q$-exponential function defined by
\begin{eqnarray}
{\rm exp}_q(x)&=&[1+(1-q)x]^{\frac{1}{1-q}},
\hspace{1.0cm}\mbox{for $1+(1-q)x > 0$} \nonumber \\
&=& 0. 
\hspace{4.0cm}\mbox{otherwise}
\end{eqnarray}
In Eqs. (1)-(6), $q$ expresses the entropic index, 
$f^B(\beta)$ the $\Gamma$ (or $\chi^2$) distribution 
function of the order $N$, 
$E(Q)$ the expectation value of $Q$
averaged over $f(\beta)$,
$\beta_0$ the average of the fluctuating $\beta$
and $2/N$ its variances.
The $\Gamma$ distribution of the order $N$
is emerging from the sum
of squares of $N$ Gaussian random variables.
In deriving Eqs. (1)-(5), we have assumed that
$N$ particles are confined within 
a small volume of $L^3$ ($L < \xi$)
where the variable $\beta$ uniformly fluctuates,
$\xi$ standing for the {\it coherence} length
\cite{Beck02}.

The important consequence of the NES is
that energy and  entropy
are not proportional to $N$ in nanosystems.
The non-extensivity of the entropy was first demonstrated
by Tsallis, who proposed the generalized entropy 
given by \cite{Tsallis88}
\begin{equation}
S_q= k_B \left( \frac{\sum_i p_i^q-1}{1-q} \right)
= - k_B \sum_i \:p_i^q \:{\rm ln}_q(p_i),
\end{equation}
where 
$p_i$ [$=p(\epsilon_i)$] denotes the probability 
distribution for the energy $\epsilon_i$ in the system 
and ${\rm ln}_q(x)$ [$=(x^{1-q}-1)/(1-q)$] the $q$-logarithmic 
function, the inverse of the $q$-exponential function
defined by Eq. (6).
It is noted that in the limit of $q = 1$,
Eq. (7) reduces to
the entropy of BGS, $S_{BG}$, given by 
\begin{equation}
S_1 = S_{BG}= - k_{B} \sum_i p_i \;{\rm ln} \:p_i.
\end{equation}
The non-extensivity in the Tsallis entropy is satisfied as follows.
Suppose that the total system containing $2N$ 
particles is divided into two 
independent subsystems, each of which contains $N$ particles,
with the probability distributions,
$p_i^{(1)}$ and $p_i^{(2)}$.
The total system is described by the factorized
probability distribution $p_{ij}=p_i^{(1)}\:p_j^{(2)}$.
The entropy for the total system $S(2N)$ is given by
\cite{Tsallis88}
\begin{equation}
S(2N)=  S(N) +S(N) +O\left( \frac{1}{N} \right),
\end{equation}
where $S(N)$ stands for the entropy of the $N$-particle subsystem,
the index $q$ given by Eq. (2) being employed.
Similarly the energy of the total system
is expressed by
\begin{equation}
E(2N)=  E(N) +E(N) +O\left( \frac{1}{N} \right),
\end{equation}
The difference of $E(2N) - 2 E(N)$ is attributed to
the surface contribution.
This implies that the index $\gamma$ in $Q \: \propto \: N^{\gamma}$
is neither 0 nor 1 for
$Q$ = $S$ and $E$ in nanosystems within the NES.

The functional form of the probability distribution
$p(E)$ expressed by Eq. (1) was originally
derived by the maximum-entropy method 
\cite{Tsallis88}\cite{Tsallis98}. 
The probability of $p_i \:[=p(\epsilon_i)]$
for the eigenvalue $\epsilon_i$ in the NES is determined 
by imposing the variational condition to
the entropy given by Eq. (7) with the two constraints
\cite{Tsallis98}:
\begin{eqnarray}
\sum_i p_i &=&1, \\
\frac{\sum_i p_i^q \epsilon_i}
{\sum_i p_i^q} &=&E_q.
\end{eqnarray}
The maximum-entropy method leads to 
the probability distribution $p_i$ given by
\begin{eqnarray}
p_i &\propto& {\rm exp}_q \left[- \beta_0 
\left( \epsilon_i-E_q \right) \right],
\end{eqnarray}
with
\begin{eqnarray}
\beta_0 &=& \frac{\beta}{c_q}, \\
c_q&=&\sum_i p_i^q,
\end{eqnarray}
where $\beta$ denotes the Lagrange multiplier
relevant to the constraint given by Eq. (12).
It has been shown that
the physical temperature $T$ of the nanosystem
is given by \cite{Abe01}
\begin{eqnarray}
T&=& \frac{c_q}{k_B \beta},
\hspace{1cm}\mbox{(AMP)}
\end{eqnarray}
In the limit of $q=1$, we get ${\rm exp}_q[x]=e^x$, $c_q=1$
and $p_i$ given by Eqs. (13)-(16)
reduces to the results obtained in the BGS,
related discussions being given in Sec. 4.

In previous papers \cite{Hasegawa04,Hasegawa05,Hasegawa05b},
I have applied the NES to the Hubbard model, which is
one of the most important models in solid-state physics
(for a recent review, see Ref. \cite{Kakehashi04}).
The Hubbard model consists of the tight-binding term
expressing electron hoppings and the short-range interaction
between two electrons with opposite spins.
The Hubbard model provides us with good
qualitative description for many interesting phenomena
such as magnetism, electron correlation, and superconductivity.
In particular, the Hubbard model has been widely employed 
for a study on transition-metal magnetism. 
In the limit of strong interaction ($U/t \ll 1$),
the Hubbard model with the half-filled electron occupancy
reduces to the Heisenberg or Ising model.
The two-site Hubbard model has been adopted for a study on 
some charge-transfer salts
like tetracyanoquinodimethan (TCNQ)
with dimerized structures \cite{Suezaki72}-\cite{Bernstein74}.
Their susceptibility and specific heat 
were analyzed by taking into account
the interdimer hopping within the BGS.
The NES calculations have been made for
thermodynamical properties of canonical 
\cite{Hasegawa04}\cite{Hasegawa05b}
and grand-canonical ensembles \cite{Hasegawa05}
of {\it Hubbard dimers}, each of which is
described by the two-site Hubbard model.
It has been shown that the temperature dependences of 
the specific heat and susceptibility is
significantly different from those calculated
by the BGS when the entropic index $q$ 
departs from unity for small $N$ [Eq. (2)],
the NES in the limit of $q=1$ reducing to the BGS.

The purpose of the present paper is 
to show 
(1) how thermodynamical property
of a nanocluster containing a small number
of Hubbard dimers is different from that
of macroscopic systems, and
(2) how thermodynamical property
of a given nanocluster is changed when $M$, the number of
Hubbard dimers contained in it, is varied.
The paper is organized as follows. In Sec. 2,
I apply the NES to nanoclustes, providing 
expressions for the energy, entropy,
magnetization, specific heat and susceptibility.
Numerical calculations of the temperature
and magnetic-field dependences of thermodynamical quantities
are reported for various $M$ values.
The final Sec. 4 is devoted to discussions and conclusions.  
In the Appendix, the NES has been applied 
to a cluster containing spin dimers
described by the Heisenberg model.

\section{Nonextensive thermodynamics of Hubbard dimers}

\subsection{Energy and entropy}

I have adopted a system
consisting of sparsely distributed $N_c$ nanoclusters,
each of which contains independent $M$ dimers.
It has been assumed that
the distance between nanoclusters is larger than
$\xi$, the coherence length of the fluctuating $\beta$ field,
and that the linear size of the clusters
is smaller than $\xi$.
Physical quantities such as the entropy and
energy are extensive for $N_c$,
but not for $M$ in general \cite{Beck02}.

The Humiltonian of the cluster is given by
\begin{eqnarray}
H &=& \sum_{\ell=1}^M \: H_{\ell}^{(d)}, \\
H_{\ell}^{(d)} &=& -t \sum_{\sigma}
( a_{1\sigma}^{\dagger} a_{2\sigma} 
+  a_{2\sigma}^{\dagger} a_{1\sigma}) 
+ U \sum_{j=1}^2 n_{j \uparrow} n_{j \downarrow }
- \mu_B B \sum_{j=1}^2 (n_{j\uparrow} - n_{j \downarrow}), \nonumber\\
&& \hspace{10cm} \mbox{($1, 2 \in \ell$)}
\end{eqnarray}
where $H_{\ell}^{(d)}$ denotes the 
two-site Hamiltonian for the $\ell$th dimer,
$n_{j\sigma}
= a_{j\sigma}^{\dagger} a_{j\sigma}$,
$a_{j\sigma}$ the annihilation operator of an electron with
spin $\sigma$ on a site $j$ ($\in \ell$), 
$t$ the hopping integral,
$U$ the intraatomic interaction, 
$\mu_B$ the Bohr magneton,
and $B$ an applied magnetic field.
In the case of the half-filled occupancy,
in which the number of electrons is $N_e=2$,
six eigenvalues of $H_{\ell}^{(d)}$ are given by
\begin{equation}
\epsilon_{i\ell}=0, \; 2\mu_B B, \; -2\mu_B B, \; U, 
\; \frac{U}{2}+\Delta, \; \frac{U}{2}-\Delta,
\hspace{1cm}\mbox{for $i=1-6,\; \ell=1-M$}
\end{equation}
where $\Delta=\sqrt{U^2/4+4 t^2}$ 
\cite{Suezaki72}\cite{Bernstein74}.
The number of eigenvalues of the total Hamiltonian
$H$ is $6^M$.

First we employ the BGS, in which
the canonical partition function for $H$ 
is given by \cite{Suezaki72}\cite{Bernstein74}
\begin{eqnarray}
Z_{BG}&=& {\rm Tr} \: {\rm exp}(-\beta H ), \\
&=& \sum_{i_1=1}^6 \cdot \cdot \sum_{i_M=1}^6
{\rm exp}[-\beta (\epsilon_{i_1} + \cdot \cdot+ \epsilon_{i_M})], \\
&=& [Z_{BG}^{(d)}]^M, \\
Z_{BG}^{(d)}&=&1+2 \:{\rm cosh}(2 \beta \mu_B B) + e^{- \beta U}
+ 2 \:e^{-\beta U/2} {\rm cosh} (\beta \Delta),
\end{eqnarray}
where $\beta=1/k_B T$,
Tr denotes the trace and $Z_{BG}^{(d)}$ the partition function
for a single dimer.
By using the standard method in the BGS, 
we can obtain various thermodynamical quantities of the system 
\cite{Suezaki72,Shiba72,Bernstein74}.
Because of a power expression given by Eq. (22),
the energy and entropy are proportional to $M$:
$E_{BG}= M E_{BG}^{(d)}$ and $S_{BG}= M S_{BG}^{(d)}$
where $E_{BG}^{(d)}$ and $S_{BG}^{(d)}$ are
for a single dimer.
This is not the case in the NES as will be discussed below.

Next we adopt the NES, where
the entropy $S_q$ for the quantum system
is defined by \cite{Tsallis88}\cite{Tsallis98}
\begin{equation}
S_q=k_B \left( \frac{{\rm Tr} \:(\rho_q^q) - 1}{1-q} \right).
\end{equation}
Here $\rho_q$ stands for 
the generalized canonical density matrix,
whose explicit form will be determined shortly [Eq. (27)].
We will impose the two constraints given by 
\begin{eqnarray}
{\rm Tr} \:(\rho_q)&=&1, \\
\frac{{\rm Tr} \:(\rho_q^q H)}{{\rm Tr} \:(\rho_q^q)}
&\equiv& <H>_q = E_q,
\end{eqnarray}
where the normalized formalism is adopted \cite{Tsallis98}.  
The variational condition for the entropy with
the two constraints given by Eqs. (25) and (26)
yields
\begin{equation}
\rho_q=\frac{1}{X_q} {\rm exp}_q 
\left[ -\left( \frac{\beta}{c_q} \right) (H-E_q) \right],
\end{equation}
with
\begin{equation}
X_q={\rm Tr}\: \left( {\rm exp}_q 
\left[-\left( \frac{\beta}{c_q} \right) (H-E_q)\right] \right),
\end{equation}
\begin{equation}
c_q= {\rm Tr} \:(\rho_q^q) = X_q^{1-q},
\end{equation}
where 
${\rm exp}_q (x)$
is the $q$-exponential function given by Eq. (6) 
and $\beta$ is a Lagrange multiplier given by
\begin{equation}
\beta=\frac{\partial S_q}{\partial E_q}.
\end{equation}
The trace in Eq. (28) and (29) is performed over
the $6^M$ eigenvalues, for example, as
\begin{eqnarray}
X_q &=&\sum_{i_1=1}^6 \cdot \cdot \sum_{i_M=1}^6
\left( {\rm exp}_q \left[ - \left( \frac{\beta}{c_q} \right)
(\epsilon_{i_1}+ \cdot \cdot + \epsilon_{i_M}-E_q)\right] \right),\\
&\ \equiv & \sum_i 
\left( {\rm exp}_q \left[ - \left( \frac{\beta}{c_q} \right)
(\epsilon_i-E_q) \right] \right),
\end{eqnarray}
where the following conventions are adopted:
\begin{eqnarray}
\epsilon_i &=& \epsilon_{i_1}+ \cdot \cdot + \epsilon_{i_M},\\
\sum_i &=& \sum_{i_1=1}^6 \cdot \cdot \sum_{i_M=1}^6.
\end{eqnarray}

It is noted that in the limit of $q = 1$,
Eq. (31) reduces to
\begin{equation}
X_1 = Z_{BG} \:{\rm exp}[\beta E_1] 
=[Z_{BG}^{(d)} \: {\rm exp}\;(\beta E_{BG}^{(d)})]^M.
\end{equation}
For $q \neq 1$, however, $X_q$ cannot be expressed
as a power form because
of the property of the $q$-exponential function:
\begin{eqnarray}
{\rm exp}_q(x+y) 
&\neq & {\rm exp}_q(x)\;{\rm exp}_q(y).
\hspace{1cm}\mbox{(for $q \neq 1$)}
\end{eqnarray}

It is necessary to point out that $E_q$ in Eq. (26) includes
$X_q$ which is expressed by $E_q$ in Eq. (28).
Then $E_q$ and $X_q$ have to be determined self-consistently
by Eqs. (26)-(29) with the $T-\beta$ relation 
given by Eq. (16) for a given temperature $T$.
The calculation of thermodynamical quantities
in the NES generally
becomes more difficult than that in BGS.

\subsection{Specific heat}

The specific heat
in the NES is given by \cite{Hasegawa04}
\begin{equation}
C_q= \left( \frac{d \beta}{d T} \right) 
\left( \frac {d E_q}{d \beta} \right).
\end{equation}
Because $E_q$ and $X_q$ are determined by
Eqs. (26)-(29), we get simultaneous equations for
$d E_q/d \beta$ and  $d X_q/d \beta$, given by
\begin{eqnarray}
\frac {d E_q}{d \beta} 
&=& a_{11}  \left( \frac {d E_q}{d \beta} \right)
+ a_{12} \left( \frac {d X_q}{d \beta} \right) + b_1, \\
\frac {d X_q}{d \beta} 
&=& a_{21}  \left( \frac {d E_q}{d \beta} \right)
+ a_{22} \left( \frac {d X_q}{d \beta} \right),
\end{eqnarray}
with
\begin{eqnarray}
a_{11}&=& q \beta X_q^{q-2} 
\sum_i w_i^{2q-1} \epsilon_i,  \\
a_{12}&=& -X_q^{-1} E_q
-\beta q (q-1) X_q^{q-3} \sum_i w_i^{2q-1}
\epsilon_i (\epsilon_i -E_q), \\
a_{21}&=& \beta X_q^q,\\
a_{22}&=& 0,\\
b_1&=& - q X_q^{q-2} \sum_i w_i^{2q-1}
\epsilon_i (\epsilon_i-E_q),\\
w_i 
&=& {\rm exp}_q \left[ - \left(\frac{\beta}{c_q} \right)
(\epsilon_i - E_q) \right],\\
%
X_q &=& \sum_i w_i.
\end{eqnarray}
The specific heat is then given by
\begin{equation}
C_q= \left( \frac{d \beta}{d T} \right) 
\left( \frac{b_1}{1-a_{11}-a_{12}a_{21}} \right).
\end{equation}
with
\begin{eqnarray}
\frac{d \beta}{d T}
&=& - \left( \frac{\beta^2}
{X_q^{1-q} - \beta (1-q) X_q^{-q} 
(d X_q/d \beta) }\right), 
%
%
\end{eqnarray}

In the limit of $q \rightarrow 1$,
Eqs. (38)-(46) yield the specific heat
in the BGS, given by \cite{Hasegawa05}
\begin{equation}
C_{BG} = \frac{d E_{BG}}{d T}
= k_B \beta^2 (<\epsilon_i^2>_1 - <\epsilon_i>_1^2),
\end{equation}
where $<\cdot>_1$ is defined by Eq. (26) with $q=1$:
\begin{equation}
< Q_i >_1 = X_1^{-1} \sum_i 
{\rm exp}[-\beta (\epsilon_i-E_1)]\:Q_i
=Z_{BG}^{-1} \sum_i {\rm exp}(-\beta \epsilon_i) \:Q_i.
\end{equation}

\subsection{Magnetization}

The field-dependent magnetization $m_q$ in the NES
is given by \cite{Hasegawa04}
\begin{eqnarray}
m_q&=& -\frac{\partial E_q}{\partial B} 
+ (k_B \:\beta)^{-1} \frac{\partial S_q}{\partial B},\\
&=& -\frac{\partial E_q}{\partial B} 
+ \beta^{-1} X_q^{-q} \frac{\partial X_q}{\partial B}.
\end{eqnarray}
By using Eqs. (26)-(29),
we get the simultaneous equations for 
$\partial E_q/\partial B$ and 
$\partial X_q/\partial B$ given by
\begin{eqnarray}
\frac{\partial E_q}{\partial B}
&=&a_{11} \frac{\partial E_q}{\partial B}
+ a_{12} \frac{\partial X_q}{\partial B} + d_1, \\
\frac{\partial X_q}{\partial B}
&=&a_{21} \frac{\partial E_q}{\partial B}
+ a_{22} \frac{\partial X_q}{\partial B} + d_2, 
\end{eqnarray}
with
\begin{eqnarray}
d_1&=& - X_q^{-1} \sum_i w_i^q \mu_i
+ \beta q X_q^{q-2} \sum_i w_i^{2q-1} \epsilon_i
\mu_i,\\
d_2&=& \beta X_q^{q-1} \sum_i w_i^q \mu_i,
\end{eqnarray}
where $\mu_i=- \partial \epsilon_i/\partial B$, and
$a_{ij}$ ($i,j=1,2$) are given by Eqs. (40)-(43).
From Eqs. (51)-(56), we obtain $m_q$ given by
\begin{eqnarray}
m_q&=&\left( \frac{-c_{12}+\beta^{-1}X_q^{-q}(1-c_{11})}
{1-c_{11}-c_{12}c_{21}} \right) d_2, \\
&=& X_q^{-1} \sum_i \:w_i^q \:\mu_i
= <\mu_i>_q. 
\end{eqnarray}

In the limit of $q \rightarrow 1$,
Eqs. (55) and (56) reduce to
\begin{eqnarray}
d_1&=& - \left< \mu_i \right>_1
+ \beta \left< \epsilon_i \mu_i \right>_1,\\
d_2&=& \beta X_1 \left< \mu_i \right>_1,
\end{eqnarray}
where $<\cdot>_1$ is given by Eq. (50).
By using Eq. (58), we get 
\begin{eqnarray}
m_{BG}&=& \left < \mu_i \right>_1,\\
&=& \frac{4 \mu_B \:{\rm sinh}(2\beta B)}{Z_{BG}},
\end{eqnarray}
where $Z_{BG}$ and $<\cdot >_1$ are
given by Eqs. (20) and (50), respectively.

\subsection{Susceptibility}

The high-field susceptibility 
in the NES is given by
\begin{equation}
\chi_q(B)= \frac{\partial m_q}{\partial B}.
\end{equation}
The zero-field susceptibility 
$\chi_q(B=0)$ is given by
\cite{Hasegawa04}
\begin{eqnarray}
\chi_q&=& \chi_q(B=0)
= -E_q^{(2)} + \beta^{-1} X_q^{-q}X_q^{(2)},
\end{eqnarray}
where 
$E_q^{(2)}=\partial^2 E_q/\partial B^2\mid_{B=0}$ and 
$X_q^{(2)}=\partial^2 X_q/\partial B^2\mid_{B=0}$.
With the use of Eqs. (26)-(29), we get simultaneous equations
for $E_q^{(2)}$ and $X_q^{(2)}$ given by
\begin{eqnarray}
E_q^{(2)}&=&a_{11} E_q^{(2)}+ a_{12} X_q^{(2)} + f_1, \\
X_q^{(2)}&=&a_{21} E_q^{(2)}+ a_{22} X_q^{(2)} + f_2, 
\end{eqnarray}
with
\begin{eqnarray}
f_1&=& -2 \:\beta \:q \:X_q^{q-2} \sum_i w_i^{2q-1}\: \mu_i^2,\\
f_2&=& \beta^2 \:q \:X_q^{2(q-1)} \sum_i w_i^{2q-1} \:\mu_i^2,
\end{eqnarray}
where $a_{ij}$ ($i,j=1,2$) are given by Eqs. (40)-(43).
From Eqs. (64)-(68), we get
\begin{eqnarray}
\chi_q&=& \frac{f_2}{a_{21}}
=\beta q X_q^{q-2} \sum_i w_i^{2q-1} 
\mu_i^2\mid_{B=0}.
\end{eqnarray}

In the limit of $q=1$, Eq. (69) yields
the susceptibility in BGS:
\begin{eqnarray}
\chi_{BG} &=& \beta
<\mu_i^2 \mid_{B=0}>_1 ,\\
&=& \left(\frac{\mu_B^2}{k_B T}\right)
\frac{8}{3+e^{-\beta U}+2 e ^{-\beta U/2}
\:{\rm cosh}(\beta \Delta)}.
\end{eqnarray}

\section{Calculated results}

\subsection{Temperature dependence}

In order to study how thermodynamical quantities of a cluster
containing $M$ Hubbard dimers depend on $M$, 
I have made some NES calculations,
assuming the $M-q$ relation given by
\begin{equation}
q =1+\frac{1}{M},
\end{equation}
which is derived from Eq. (2) with $M=2 N$
for dimers.
Simultaneous equations for $E_q$ and $X_q$
given by Eqs. (26)-(29) have been solved
by using the Newton-Raphson method 
with initial values of $E_1$ and $X_1$
obtained from BGS ($q=1$) corresponding
to $M=\infty$ in Eq. (72). 
Calculated quantities are given {\it per dimer}.

Figures 1(a), 1(b) and 1(c) show the temperature
dependence of the specific heat $C_q$ 
for $U/t=0$, 5 and 10, respectively, with
various $M$ values.
The specific heat for $M=\infty$
shown by bold solid curves, expresses
the result in BGS, and it 
has a peak at lower temperatures for the larger interaction,
as previous BGS calculations showed \cite{Shiba72}.
Note that the horizontal scales of Fig. 1(c) are
enlarged compared to those of Figs. 1(a) and 1(b).
The peak becomes broader for smaller $M$.

The temperature dependence of the susceptibility $\chi_q$
for $U/t=0$, 5 and 10 is plotted 
in Figs. 2(a), 2(b) and 2(c), respectively.
The susceptibility for $M=\infty$ (BGS) 
shown by the bold solid curve, has 
a larger peak at lower temperatures for
larger $U$ \cite{Shiba72}.
Note that the horizontal and vertical
scales of Fig. 2(c) are different from those
of Figs. 2(a) and 2(b).
We note that for smaller $M$, the peak
in $\chi_q$ becomes broader, which is similar to
the behavior of the specific heat shown in Figs. 1(a)-1(c).

When the $M$ value is varied,
maximum values of the specific heat
($C^*_q$) and the susceptibility ($\chi^*_q$) are changed, 
and the temperatures ($T^*_C$ and $T^*_{\chi}$) 
where these maxima are realized, are also changed.
Figure 3(a) depicts $T^*_C$ and $T^*_{\chi}$ for $U/t=5$
as a function of $1/M$. 
It is shown that with increasing $1/M$,
$T^*_{\chi}$ is much increased than $T_C^*$.
Similarly, the $1/M$ dependences of
$C^*_q$ and $\chi^*_q$ for $U/t=5$ are plotted in
Fig. 3(b), which shows that maximum values of $C_q$
and $\chi_q$ are decreased with decreasing $M$.
This trend against $1/M$ is due to the fact that a decrease 
in $M\:(=2 N)$
yields an increase in fluctuations of $\beta$ fields, 
and then peaked structures of the specific
heat and susceptibility realized in the BGS, 
are smeared out by $\beta$ in Eq. (1).

\subsection{Magnetic-field dependence}

Next I discuss the magnetic-field dependence
of physical quantities.
Figure 4 shows the $B$ dependence of the magnetization $m_q$ 
for $U/t=0$, 5 and 10 with $M=2$ at $k_B T/t=1$.
For $U/t=0$, $m_q$ in the NES is smaller
than that in the BGS at $\mu_B B/t < 1$, but at $\mu_B B/t > 1$
the former becomes larger than the latter.
In contrast, in cases of $U/t=5$ and 10, $m_q$ in the
NES is larger than that in the BGS for $\mu_B B/t > 0$.
In order to study the $B$ dependence in more details, 
I show in Fig. 5 the $B$ dependence of the six eigenvalues
of $\epsilon_i$ for $U/t=5$ [Eq. (19)].
We note the crossing of the lowest eigenvalues
of $\epsilon_3$ and $\epsilon_6$ at the critical filed:
\begin{equation}
\mu_B B_c = \sqrt{\frac{U^2}{16}+ t^2}-\frac{U}{4},
\end{equation}
leading to $\mu_B B_c/t=0.351$ for $U/t=5.0$.
At $B < B_c$ ($B > B_c$),
$\epsilon_6$ ($\epsilon_3$) is the ground state.
At $B=B_c$ the magnetization $m_q$ is rapidly increased
as shown in Figs. 6(a) and 6(b) for $k_B T/t=1.0$
and 0.1, respectively:
the transition at lower temperatures is more
evident than at higher temperatures. 
This level crossing also yields a peak in $\chi_q$ 
[Figs. 6(c) and 6(d)] and
a dip in $C_q$ [Figs. 6(e) and 6(f).
It is interesting that
the peak of $\chi_q$ for $M=2$ 
is more significant than that
for $M=\infty$ whereas the dip of $C_q$ for $M=2$
is broader than that for $M=\infty$. 
When the temperature becomes higher, these peak structures
become less evident as expected.
Similar phenomenon in the field-dependent specific heat
and susceptibility have been pointed out in the
Heisenberg model within the BGS \cite{Kuzmenko04}.

In the case of the quarter-filled occupancy ($N_e=1$), 
the eigenvalues
are $\epsilon_i=-t-\mu_B B$, $-t+\mu_B B$,
$t-\mu_B B$, and $t+\mu_B B$ for $i=1-4$.
Although the level crossing occurs between
$\epsilon_2$ and $\epsilon_3$ at $\mu_B B=t$,
it does not show any interesting behavior
because the crossing occurs between 
the excited states.
The case for the three-quarter-filled occupancy ($N_e=3$)
is the same as that of the quarter-filled occupancy
because of the electron-hole symmetry of the model. 

Figure 6(b) reminds us the quantum tunneling of magnetization
observed in magnetic molecular clusters such as Mn12 and Fe8
\cite{Mn12}, which originates from the level crossings of 
magnetic molecules when a magnetic field is applied \cite{Mn12}.

\section{Discussions and conclusions}

I have applied the NES to Hubbard dimers for a study
of their thermodynamical properties.
The current NES is, however, still in its infancy, having
following unsettled issues.

\noindent
(i) For relating the physical temperature $T$
to the Lagrange multiplier $\beta$, 
I have employed the $T-\beta$ relation
given by Eq. (16).
There is an alternative proposal with 
the $T-\beta$ relation given by\cite{Tsallis98}
\begin{eqnarray}
T&=&\frac{1}{k_B \beta},
\hspace{1cm}\mbox{(TMP)}
\end{eqnarray}
which is the same as in the BGS.
At the moment, it has not been established 
which of the AMP and TMP methods given by Eqs. (16) and (74),
respectively,
is appropriate as the $T-\beta$ relation in the current NES.
It has been demonstrated that the negative specific heat
of a classical gas model realized
in the TMP method \cite{Abe99}, 
is remedied in the AMP method \cite{Abe01}.
Recent theoretical analyses also suggest that
the AMP method is better than the TMP method
\cite{Suyari05}\cite{Wada05}.
The TMP method yields an anomalously large
Curie constant of the susceptibility
in the free spin model \cite{Hasegawa05}\cite{Mar00}
and in the Hubbard model \cite{Hasegawa04,Hasegawa05}.
In my previous papers \cite{Hasegawa04}-\cite{Hasegawa05b}, 
NES calculations have been made by using
the TMP and AMP methods. It has been
shown that both methods
yield qualitatively similar results although
there are some quantitative difference between the two:
the non-extensivity in the TMP method generally appears more
significant than that in the AMP method.

\noindent
(ii) The $N-q$ relation given by Eq. (2) 
was obtained in Eqs. (1)-(5) with the $\Gamma$ distribution
$f^B(\beta)$ given by Eq. (3).
Alternatively, by using the large-deviation approximation,
Touchette \cite{Touchette02} has obtained the distribution
function $f^T(\beta)$, in place of $f^B(\beta)$, given by
\begin{eqnarray}
f^T(\beta)&=& \frac{\beta_0}{\Gamma 
\left( \frac{N}{2} \right)}
\left( \frac{N \beta_0}{2} \right)^{\frac{N}{2}}
\beta^{-\frac{N}{2}-2} 
{\rm exp}\left( -\frac{N \beta_0}{2\beta} \right).
\end{eqnarray}
For $N \rightarrow \infty$,
both $f^B(\beta)$ and $f^T(\beta)$ distribution functions 
reduce to the delta-function densities, and
for a large $N \; (> 100)$, both distribution functions
lead to similar results.
For a small $N \;(< 10)$, however, 
there is a clear difference between the two distribution 
functions (see Fig. 4 of Ref.\cite{Hasegawa05b}).
It should be noted that $f^T$ cannot lead to the
$q$-exponential function which plays a crucial role
in the NES. For a large $\epsilon$,
the $\Gamma$ distribution $f^B$ in Eq. (1) yields the power form of
$w(\epsilon) \sim \epsilon^{-\frac{1}{q-1}}$ while
$f^T$ substituted to Eq. (1)
leads to the stretched exponential form of
$w(\epsilon) \sim e^{c \sqrt{\epsilon}}$.
This issue of $f$ versus $f^T$ is related 
to the {\it superstatistics},
which is currently studied with much interest
\cite{Beck04}.

To summarize,
I have discussed thermodynamical properties
of a nanocluster containing $M$ dimers,
applying the NES to
the Hubbard model.
It has been demonstrated that the thermodynamical properties
of a nanocluster with a small $M$
calculated by the NES
may be considerably different from those
obtained by the BGS.
It is interesting to compare our theoretical prediction
with experimental results for samples containing
a small number of transition-metal dimers. 
Unfortunately samples with such a small number of dimers 
have not been reported:
samples having been so far synthesized include
macroscopic numbers of dimers, to which the present
analysis cannot be applied. 
I expect that it is possible to form a dimer assembly 
by STM manipulation of individual atoms \cite{Manoharan00}.
Scanning probes may be used also as dipping pens to write 
small dimerized structures \cite{Piner99}.
Theoretical and experimental
studies on nanoclusters with changing $M$
could clarify a link between the behavior of the
low-dimensional infinite systems and finite-size nanoscale systems.
I hope that the unsettle issues (i) and (ii) in the
current NES mentioned above
are expected to be resolved by future experiments
on nanosystems with changing their sizes.
It would be interesting to adopt quantum-master-equation
and quantum-Langevin-equation approaches,
and/or to perform large-scale molecular-dynamical simulations,
for nanoclusters described by the Hubbard model.

\section*{Acknowledgements}
It is my great pleasure that on the occasion 
of the 60th birthday of Professor David G. Pettifor,
I could dedicate the present paper to him, with whom 
I had an opportunity of collaborating 
in Imperial College London 
for one year from 1980 to 1981.


\vspace{1cm}
\noindent
{\large\bf Appendix: NES for Heisenberg dimers}

I have considered a cluster containing $M$ spin dimers
(called {\it Heisenberg dimers})
described by the Heisenberg model ($s=1/2$) given by
\begin{eqnarray}
H &=& \sum_{\ell=1}^M H_{\ell}^{(d)}, \\
H_{\ell}^{(d)} &=& -J {\bf s}_1 \cdot {\bf s}_2 
- g \mu_B B (s_{1z}+s_{2z}),
\hspace{1cm}\mbox{($1,2 \in \ell$)}
\end{eqnarray}
where $J$ stands for the exchange interaction,
$g$ (=2) the g-factor, $\mu_B$ the Bohr magneton,
and $B$ an applied magnetic field.
Four eigenvalues of $H_{\ell}^{(d)}$ are given by
\begin{eqnarray}
\epsilon_{i \ell} &=& -\frac{J}{4}-g \mu_B B m_i,
\hspace{2cm}\mbox{with $m_1=1,\:0,\: -1$ for $i=1, 2, 3$},
\nonumber \\
&=& \frac{3 J}{4}-g \mu_B B m_i.
\hspace{2cm}\mbox{with $m_4=0$ for $i=4$},
\end{eqnarray}

In the BGS the canonical partition function is given by
\cite{Mentrup99}-\cite{Dai03}
\begin{eqnarray}
Z_{BG} &=& [Z_{BG}^{(d)}]^M, \\
Z_{BG}^{(d)} &=& {\rm exp} \left(\frac{\beta J}{4} \right) 
[1+2{\rm cosh} (g \mu_B \beta B)]
+{\rm exp}\left(-\frac{3\beta J}{4} \right),
\end{eqnarray}
with which thermodynamical quantities are
easily calculated.
The susceptibility is, for example, given by
\begin{eqnarray}
\chi_{BG} &=& M \chi_{BG}^{(d)}, \\
\chi_{BG}^{(d)} &=&
\frac{\mu_B^2}{k_B T} \left( \frac{8}
{3+ {\rm exp}(-J/k_B T)} \right).
\end{eqnarray}

The calculation of thermodynamical quantities
in the NES for the Heisenberg
model goes parallel
to that discussed in Sec. 2 if we employ
eigenvalues given by Eq. (78).
For example, by using Eq. (69),
we get the susceptibility for the
Heisenberg model, given by
\begin{eqnarray}
\chi_q 
&=& g^2 \mu_B^2 \left( \frac{q \beta}{c_q} \right)
\: \frac{1}{X_q} \sum_i w_i^{2q-1} m_i^2.
\end{eqnarray}
In the case of $M=1$ (a single dimer), we get
\begin{eqnarray}
\chi_q^{(d)} &=& g^2 \mu_B^2 \left( \frac{q \beta}{c_q} \right)
\left( \frac{2}{X_q} \right)
\left({\rm exp}_q \left[ \left( \frac{\beta}{c_q} \right) 
\left( \frac{J}{4}+E_q \right) \right]
\right)^{2q-1}, 
\end{eqnarray}
with
\begin{eqnarray}
X_q &=& 3 \:{\rm exp}_q
\left[ \left( \frac{\beta}{c_q}\right)
\left( \frac{J}{4}+E_q \right) \right]
+{\rm exp}_q
\left[\left( -\frac{\beta}{c_q} \right)
\left( \frac{3J}{4}-E_q \right) \right], \\
E_q &=& \frac{1}{X_q} \{
\left(\frac{-3J}{4} \right) 
\left( {\rm exp}_q \left[ \left( \frac{\beta}{c_q} \right)
\left( \frac{J}{4}+E_q \right) \right] \right)^q \nonumber \\
&+& \left( \frac{3J}{4} \right)
\left( {\rm exp}_q \left[\left(-\frac{\beta}{c_q} \right)
\left (\frac{3J}{4}-E_q \right) \right] \right)^q 
\}.
\end{eqnarray}
In the limit of $q=1$,
Eq. (84) reduces to $\chi_{BG}^{(d)}$ given by Eq. (82).

The Curie constant $\Gamma_q$ defined by
$\chi_q=(\mu_B^2/k_B)(\Gamma_q/T)$ for $T \gg J$ 
is given by
\begin{eqnarray}
\Gamma_q 
&=& 2 M \: q,
\hspace{4cm}\mbox{(AMP)} \\ 
&=& 2 M \:q \:4^{M(q-1)}. 
\hspace{2cm}\mbox{(TMP)} 
\end{eqnarray}
Equations (87) and (88) are derived with the use of
the $T-\beta$ relation given by Eqs. (16) and (74),
respectively.
These are consistent with results obtained 
for Hubbard dimes \cite{Hasegawa05}.

Figures 7(a) and 7(b) show the temperature
dependence of the specific heat $C_q$ and 
susceptibility $\chi_q$ of Heisenberg dimers calculated
with the use of Eq. (83) for $M=$ 1, 2, 3 and $\infty$ 
($M=\infty$ corresponding to the BGS with $q=1.0$).
We note that the results of Heisenberg dimers are
quite similar to those of the Hubbard dimer 
for $U/t=5$ and 10 shown in Figs. 2(b), 2(c), 2(e) and 2(f).
This is not surprising because the Hubbard model 
with the half-filled electron occupancy in the
strong-coupling limit reduces to the Heisenberg model.



\newpage

\begin{figure}
\caption{
The temperature dependence of the specific heat $C_q$
{\it per dimer}
for (a) $U/t=0$, (b) 5 and (c) 10,
calculated for
$M=1$ (solid curves), 2 (chain curves),
3 (dashed curves) and 
$\infty$ (bold solid curves),
results for $M=\infty$ denoting those in the BGS.
}
\label{fig1}
\end{figure}

\begin{figure}
\caption{
The temperature dependence of the susceptibility $\chi_q$
{\it per dimer}
for (a) $U/t=0$, (b) 5 and (c) 10,
calculated for
$M=1$ (solid curves), 2 (chain curves),
3 (dashed curves) and 
$\infty$ (bold solid curves),
results for $M=\infty$ denoting those in the BGS.
}
\label{fig2}
\end{figure}

\begin{figure}
\caption{
(a) $1/M$ dependence of the temperatures of 
$T^*_C$ (circles) and $T^*_{\chi}$ (squares) where 
$C_q$ and $\chi_q$ have the maximum values, respectively.
(b) $1/M$ dependence of the maximum values of 
$C^*_q$ (circles) and $\chi^*_q$ (squares)
($U/t=5$)
}
\label{fig3}
\end{figure}

\begin{figure}
\caption{
The magnetic-filed dependence of the magnetization
$m_q$ for (a) $U/t=0$, (b) 5, and (c) 10
with $k_B T/t=1$ for $M=2$ (solid curves)
and $\infty$ (dashed curves).
}
\label{fig4}
\end{figure}

\begin{figure}
\caption{
The magnetic-filed dependence of
the eigenvalues $\epsilon_i$ ($i = 1-6$)
for $U/t=5$, $B_c$ denoting the critical field
where a level crossing
between $\epsilon_3$ and $\epsilon_6$ occurs.
}
\label{fig5}
\end{figure}

\begin{figure}
\caption{
The magnetic-filed dependence of
(a) the magnetization $m_q$ 
for $k_B T/t=1.0$ and (b) $k_B T/t=0. 1$,
(c) the susceptibility $\chi_q$
for $k_B T/t=1.0$ and (d) $k_B T/t=0. 1$, and
(e) the specific heat $C_q$ 
for $k_B T/t=1.0$ and (f) $k_B T/t=0. 1$
with U/t=5, 
calculated for $M=2$ (solid curves) 
and $\infty$ (dashed curves).
}
\label{fig6}
\end{figure}

\begin{figure}
\caption{
The temperature dependence of (a) the specific heat 
and (b) susceptibility of Heisenberg dimers 
for various $M$: $M=1$ (bold solid curves),
$2$ (chain curves), $3$ (dashed curves),
and $\infty$ (solid curves).
}
\label{fig7}
\end{figure}


\begin{thebibliography}{99}

\bibitem{Bader02}Bader SD.
Surf. Sci. 2002;500:172.

\bibitem{Kach03}Kachkachi H, Garanin DA. 
e-print: cond/mat/0310694.

\bibitem{Luban04}Luban M. 
J. Magn. Magn. Mat. 2004;272-276:e635.

\bibitem{Heer90}de Heer WA, Milani P, Ch\'{a}telain A.
Phys. Rev. Lett. 1990;65:488.

\bibitem{Bucher91}Bucher JP, Douglass DC, Bloomfield LA.
Phys. Rev. Lett. 1991;66:3052.

\bibitem{Apsel96}Aspel SE, Emmert JW, Deng J,
Bloomfield LA.
Phys. Rev. Lett. 1996;76:1441.

\bibitem{Gambardella02}Gambardella P, Dallmeyer A, Malti M,
Malagoll MC, Eberhardt W, Kern K, Carbone C.
Nature 2002;416:301.

\bibitem{Yamamoto04}Yamamoto Y, Miura T, Suzuki H, Kawamura N,
Nakamura T, Kobayashi K, Teranishi T, Hori H.
Phys. Rev. Lett. 2002;93:116801.

\bibitem{Shinohara03}Shinohara T, Sato T,
Phys. Rev. Lett. 2003;91:197201.

\bibitem{Postnikov04}Postnikov AV, Br\"{u}ger M, Schnack J.,
e-print: cond-mat/0404343.

\bibitem{Lasc97}Lascialfari A, Gatteschi D, Borsa F, Cornia A,
Phys. Rev. B 1997{\bf 55}, 14341 ().

\bibitem{Gatteshi00}Gatteshi D, Sessoli R, Cornia A.
Chem. Commun. 2000;9:725.

\bibitem{Luban02}Luban M, Borsa F, Bud'ko S, Canfield P,
Jun S, Jung JK, K\"{u}gerler P, Mentrup D, M\"{u}ller A, 
Modler R, Procissi D, Suh BJ, Torikachvili M.
Phys. Rev. B 2002;66:054407.

\bibitem{Mn12}Wernsdorfer W, Aliaga-Alcalde A,
Hendrickson DN, Christou G. \\
Nature 2002;416:406; related references therein.

\bibitem{Caciuffo98}Caciuffo R, Amoretti G, Murani A.
Phys. Rev. Lett. 1998;81:4744.

\bibitem{Fe2}Gall FL, DeBiani FF, Caneschi A,
Cinelli P, Cornia A, Fabretti AC, Gatteschi D.
Inorg. Chim. Acta 1997;262:123; \\ 
Lascialfari A, Tabak F, Abbati GL, Borsa F,
Corti M, Gatteschi D.
J. Appl. Phys. 1999;85:4539.

\bibitem{Mentrup99}Mentrup D Schnack, J, Luban M.
Physica A 1999;272:153.

\bibitem{Efremov02}Efremov DV, Klemm RA.
Phys. Rev. B 2002;66:174427; \\
cond-mat/0409168.
\bibitem{Dai03}Dai D, Whangbo M.
J. Chem. Phys. 2003;118:29.


\bibitem{V2}Furukawa Y, Iwai A, Kumagai K,
Yabubovsky A.
J. Phys. Soc. Jpn. 1996;65:2393; \\
Tennant DA, Nagler SE, Garrett AW,
Barnes T, Torardi CC.
Phys. Rev. Lett. 1997;78:4998; \\
Garrett AW, Nagler SE, Tennant DA,
Sales BC, Barnes T.
Phys. Rev. Lett. 1997;79:745.


\bibitem{Cr2}Bailey MS, Obrovac MN, Baillet E,
Reynolds TK, DiSalvo FJ.
Inorg. Chem. 2003;42:5572; \\
Glerup J, Goodson PA, Hodgson DJ,
Masood MA, Michelsen K.
Inorganica 2005;358:295.



\bibitem{Co2}Beckmann U, Brooker S.
Coordination Chemistry 2003;245:17.

\bibitem{Lazarov04}Lazarov ND, Spasojevic V, Kusigerski V,
Mati\'{c} VM, Mili\'{c} M.
J. Magn. Magn. Matt. 2004;272-276:1065.

\bibitem{Ni2}Dey SK, Fallah MSE, Ribas J,
Matsushita T, Gramlich V, Mitra S.
Inorganica Chmica 2004;357:1517.

\bibitem{Cu2}Zheludev A, Shirane G, Sasago Y, Hase M,
Uchinokura K. \\
Phys. Rev. B 1996;53:11642.

\bibitem{Hill01}Hill TL.
Nano Lett. 2001;1:273; 
{\it ibid.} 2002;2:609.

\bibitem{Jarz97}Jarzynski C.
Phys. Rev. Lett. 1997;78:2690;
Phys. Rev. E 1997;56:5018.

\bibitem{Tsallis88}Tsallis C. 
J. Stat. Phys. 1988;52:479. 

\bibitem{Tsallis98}Tsallis C, Mendes RS, Plastino AR.
Physica A 1998;261:534. 

\bibitem{Tsallis04}For a recent review on the NES, see
Tsallis C. Physica D 2004;193:3.

\bibitem{Wilk00}Wilk G, Wlodarczyk Z. 
Phys. Rev. Lett. 2000;84:2770.

\bibitem{Beck02}Beck C.  
Europhys. Lett. 2002;57:329.

\bibitem{Raja04}Rajagopal AK, Pande CS, Abe S.
eprint cond-mat/0403738.

\bibitem{Ritort04}Ritort F. e-print cond-mat/0401311.

\bibitem{Abe01}Abe S, Mart\'{i}nez S, Pennini F,
Plastino A. 
Phys. Lett. A 2001;281:126.

\bibitem{Hasegawa04}Hasegawa H.  
cond-mat/0408699.

\bibitem{Hasegawa05}Hasegawa H. 
Physica A 2005;351:273.

\bibitem{Hasegawa05b}Hasegawa H. 
Prog. Theor. Phys. suppl. 2005;XXX:YYY (in press).

\bibitem{Kakehashi04}Kakehashi Y. 
Adv. Phys. 2004;53:497;
related references therein.

\bibitem{Suezaki72}Suezaki Y.  
Phys. Lett. 1972;38A:293.

\bibitem{Shiba72}Shiba H, Pincus PA. 
Phys. Rev. B 1972;5:1966.

\bibitem{Bernstein74}Bernstein U, Pincus  P.
Phys. Rev. B 1974;10:3626.

\bibitem{Kuzmenko04}Kuzmenkoand NK, Mikhajlov  VM.
e-print cond-mat/0401468.

\bibitem{Abe99}Abe S. Phys. Lett. A 1999;263:424;
{\it ibid.} 2000;267:456(erratum).

\bibitem{Suyari05}H. Suyari, cond-mat/0502298.

\bibitem{Wada05}T. Wada and A. M. Scarfone, 
cond-mat/0502394.

\bibitem{Mar00}Martinez S, Pennini F, Plastino A. 
Physica A 2000;282:193.

\bibitem{Touchette02}Touchette H. 
e-print cond-mat/0212301.

\bibitem{Beck04}Beck C, Cohen EGD.
e-print cond-mat/0205097;
Touchette H, Beck C.
e-print cond-mat/0408091.

\bibitem{Manoharan00}Manoharan HC, Lutz CP, Eiger DM.
Nature 2000;403:512.

\bibitem{Piner99}Piner RD, Zhu J, Xu F, Hong S, Mirkin CA.
Science 1999;283:661.


\end{thebibliography}
\end{document}